\documentclass[letterpaper,12pt]{article}

\pdfoutput=1
\usepackage{hyperref}

\makeatletter
\renewcommand\section{\@startsection {section}{1}{\z@}%
{-3.5ex \@plus -1ex \@minus -0.2ex}%
{2.3ex \@plus 0.2ex}%
{\normalfont\normalsize\bfseries}}

\renewcommand\subsection{\@startsection{subsection}{2}{\z@}%
{-3.25ex \@plus -1ex \@minus -0.2ex}%
{1.5ex \@plus 0.2ex}%
{\normalfont\normalsize\bfseries}}

\def\@seccntformat#1{\csname the#1\endcsname.\quad}
\makeatother

\begin{document}

\setlength{\baselineskip}{3.6ex}

\noindent
\textbf{\LARGE A very short guide to IOI}

\vspace{2ex}
\noindent
\textbf{\large A general framework for statistical inference summarised}

\vspace{6ex}
\noindent
\textbf{Russell J. Bowater}\\
\emph{Independent researcher, Sartre 47, Acatlima, Huajuapan de Le\'{o}n, Oaxaca, C.P.\ 69004,
Mexico. Email address: as given on arXiv.org. Twitter profile:
\href{https://twitter.com/naked_statist}{@naked\_statist}\\ Personal website:
\href{https://sites.google.com/site/bowaterfospage}{sites.google.com/site/bowaterfospage}}

\vspace{5ex}
\noindent
\textbf{Abstract:}
Integrated organic inference (IOI) is discussed in a concise and informal way with the aim that the
reader is given the gist of what this approach to statistical inference is about as well as given
pointers to further reading.

\vspace{2ex}

\vspace{4ex}
\section{Introduction}
\label{sec5}

This is a brief guide to integrated organic inference or IOI for short, which is a general
framework for carrying out statistical inference.

Here, we will be concerned with the broad issue of making inferences about the parameters $\theta =
\{\theta_i:1,2,\ldots,k\}$ of a given model on the basis of an observed data set $x$ randomly
generated from this model. This issue is important both in its own right, and because tackling this
issue can be viewed as a natural stepping stone to resolving the problem of what to do when the
underlying true model is unknown.

\vspace{2ex}
\section{Objective}

We will assume that the ultimate goal of statistical inference under a given model is:

\begin{quote}
To place a probability distribution over the parameters of the model that represents what is known
about those parameters after the data have been observed. We will call this \emph{probabilistic
inference}. A method of inference that is unable to achieve this objective will be referred to as a
procedure for performing \emph{sub-probabilistic inference.}
\end{quote}

Now just before some readers run away on hearing this news, let me give you the reasons for this
being the objective:

\begin{enumerate}

\item Scientists generally want post-data uncertainty about the unknown true values of parameters
to be expressed in the form of a probability distribution over those parameters.

For example, you would not be much of a weather forecaster if rather than saying what is your
probability that there will be more than one centimetre of rain tomorrow, offered as a substitute
some kind of sub-probabilistic measure of uncertainty in relation to this event, e.g.\ a P value,
confidence interval or likelihood ratio. In this type of scenario, you would expect the people
listening to the forecast to be screaming out ``just give us a damn probability!''

\item Only in rare cases would it be better to try to persuade a scientist to accept an inference
that is sub-probabilistic.

For example, although expressing inferences in terms of imprecise probability (i.e.\ putting lower
and upper bounds on the post-data probability of an event) may be appropriate in some cases, those
cases, at best, will be few and far between. One of the reasons for this is that it is difficult to
give lower and upper probability bounds a tangible meaning, and as a result, we may often ask how
far out these limits could possibly extend.

\end{enumerate}

Given that some readers will still be sitting uncomfortably, let us ask:

\vspace{2.5ex}
\emph{Why are there strong objections to probabilistic inference being a sensible goal?}

\vspace{2.5ex}
\noindent
This is arguably due to the fact that, unfortunately, many will think that to achieve this goal you
need to be Bayesian, i.e.\ you need to plug a subjective prior distribution for the parameters
concerned into Bayes' theorem. Given their strong objection to the Bayesian method, they therefore
persuade themselves that the fruit of this method, i.e.\ probabilistic inference, itself must be
undesirable. In this sense, they are guilty of throwing the baby (probabilistic inference) out with
the bathwater (Bayes' theorem).
As will be argued, the Bayesian method is only one of a number of methods that can be used for
drawing probabilistic inferences.
However, to be able to see that clearly, you need to think about what is the true meaning of the
concept of probability and not simply take its meaning for granted.

\vspace{2ex}
\section{Statistics by analogy}
\label{sec4}

Making analogies is a fundamental part of performing statistical inference. Indeed, arguably the
most natural way of making sense of the concept of probability is through the use of analogy. Some
of these analogies may be good, some adequate and some poor. In many situations, there is a variety
of competing analogies that we can make and choosing to use one of these analogies rather than
another can radically affect what method of inference will in fact be implemented.
It should be obvious without the need for further elaboration that we should always want to make
the best analogies as possible. The mathematical tradition in the field of statistics has
overlooked the importance of the use of analogy in making inferences, which has led to much
inconsistency and many flaws in statistical theory.

To give an example, let us consider how we can make use of an analogy to interpret the concept of
probability. In fact, this interpretation seems so obvious it barely feels as though it needs to be
pointed out explicitly. Nevertheless, so many other interpretations of the meaning of probability
have been put forward, we appear to have lost our way in this respect.

In particular, let us imagine that there is an urn of balls containing a number of red balls and
yellow balls. Now let us consider the event of drawing out a red ball from the urn and also the
event of there being more than one centimetre of rain tomorrow. Here by supposing that the
proportion of red balls in the urn can be varied and by making an analogy between our confidence
that the ball drawn out of the urn will be red and our confidence that there will be more than one
centimetre of rain tomorrow, we can both determine our probability for the latter event and
interpret the meaning of this probability.

Of course, the analogy between our confidence that the ball drawn out of the urn will be red and
our confidence that any given real-world event will occur may, for example, be very strong,
moderately strong, weak or very weak which leads to the idea that once probabilities and
probability distributions have been determined they can be characterised by placing them on a scale
that reflects the strength of this type of analogy.
This generalisation of the concept of probability and the fundamental idea of defining probability
through analogies with the outcomes of standard physical experiments was explored and developed in
Bowater~(2017a) and Bowater~(2018b).

\vspace{2ex}
\section{Bayesian inference}
\label{sec1}

For many, the Bayesian method is the most obvious way of obtaining a post-data or posterior
distribution for the parameters of a given model. There are some though (actually quite a few
people!) who go further and claim it is the only way of performing this task. As a result, they are
led to try to find a Bayesian logic behind the assignment of any post-data distribution to the
parameters of a model. The view that this Bayesian logic must exist and completely justifies the
use of the post-data distribution in question is, however, a flawed view as we will now discuss.

Bayes' theorem, which provides the mathematical basis for Bayesian inference, is as follows:
\vspace{2ex}
\[
P(A\,|\,B) = \frac{P(B\,|\,A)P(A)}{P(B)}
\vspace{2ex}
\]
where $A$ and $B$ are two events. We note that if the events $A$ and $B$ are generated from the
joint distribution of these two events through the use of a standard physical experiment, e.g.\
randomly drawing balls out of an urn or spinning a wheel, then the use of Bayes' theorem is almost
without controversy and feels completely intuitive. However, this is not generally how Bayes'
theorem is used to draw inferences about the parameters of a model on the basis of an observed data
set.

So what options do we have to justify the use of Bayes' theorem in this latter scenario?

\begin{enumerate}

\item Make dubious assumptions (such as the one required to use the Dutch-book argument).

\vspace{-0.5ex}
\item Embrace dubious axioms (such as the sure-thing principle or the independence axiom).

\vspace{-0.5ex}
\item Make an analogy with the first type of scenario, i.e.\ an analogy with the outcomes of a
standard physical experiment.

\end{enumerate}

It should be obvious therefore that the third option is going to be advocated. In particular, the
analogy being referred to will be called Bayes' analogy. This terminology is indeed very
appropriate as Thomas Bayes himself used such an analogy to justify the use of Bayes' theorem to
make post-data inferences about a binomial proportion in his famous 1763 paper that originally
presented this theorem (with his choice of standard physical experiment being the random throwing
of balls onto a square table).

Since Bayes' analogy is the justification for using Bayes' theorem, if we are not prepared to make
this analogy, then we are not entitled to use Bayes' theorem. Indeed, we will take this as being
the general philosophy for all methods of inference that we may wish to advocate. In short, the
message is: \emph{no analogy, no method\hspace{0.05em}!}

Returning to the original point of discussion, if we are presented with a post-data distribution
for the parameters of a model that has been derived in some unknown manner, we may ask:

\vspace{2.5ex}
\setlength{\leftskip}{1.2em}
\noindent
\emph{Are we entitled to use Bayes' theorem in reverse to discover what was the prior distribution
for the parameters in question?}

\setlength{\leftskip}{0em}
\vspace{2.5ex}
\noindent
Here, we would of course be entitled to use Bayes' theorem in reverse if we accept the use of
Bayes' analogy in the given context. However, if the post-data distribution of the parameters
concerned was not derived using Bayes' theorem because the use of Bayes' analogy was not accepted
in that context, i.e.\ in the forward direction, why on earth should we be willing to regard the
Bayes' analogy as being acceptable in going in the reverse direction?

Therefore, we reject the idea that Bayesian inference is the only means by which a post-data
distribution for the parameters of a model can be formed. However, we maintain that Bayesian
inference will often be a convenient method for performing this task, under the condition of course
that making the Bayes' analogy is acceptable in the context of interest. This is the reason why
Bayesian inference is a key component of the IOI framework being discussed in the present guide.

\vspace{2ex}
\section{General framework of IOI: Divide and conquer!}
\label{sec2}

As just alluded to, we will not advocate using the same method of inference for all inferential
problems we may encounter. In fact, we will not necessarily advocate using just one method of
inference to tackle any specific inferential problem.

In particular, the general approach to inference (which could be referred to as an approach of
\emph{divide and conquer\hspace{0.05em}!}) will be to proceed via the following two steps:

\vspace{2.5ex}
\setlength{\leftskip}{1.2em}
\noindent
Step 1: Divide the parameter space into regions and use whatever method of inference is appropriate
in each region conditional on the parameter (which may well be a multivariate parameter) lying in
the region concerned. Therefore, for each of the regions concerned, we obtain the post-data
distribution of the parameter $\theta$ (according to the notation given in Section~\ref{sec5})
conditional on $\theta$ lying in the given region.

\vspace{2.5ex}
\noindent
Step 2: Use whatever method of inference is appropriate to determine what is the post-data
probability that the parameter $\theta$ lies in each of the regions.

\setlength{\leftskip}{0em}
\vspace{2.5ex}
In the simple case where these regions of the parameter space, which we will denote as
$\theta^{(1)}, \theta^{(2)}, \ldots, \theta^{(m)}$, are disjoint, the post-data density of the
parameter $\theta$, i.e.\ the density $p(\theta\,|\,x)$, is then determined using the formula:
\vspace{0.5ex}
\[
\textstyle
p(\theta\,|\,x) = \sum_{i=1}^{m} p(\theta\,|\,\{\theta \in \theta^{(i)}\},x)
p(\theta \in \theta^{(i)}\,|\,x)
\vspace{0.5ex}
\]
On the other hand, if the parameter space regions in question are not disjoint, it may be possible
to determine a post-data distribution of $\theta$ over the entire parameter space with\-out the
need for Step 2 of this approach.
This may be the case, for example, when we try to construct a joint post-data distribution of the
set of parameters $\theta$ by first determining the full conditional post-data distributions of
these parameters, which clearly would be a task that is carried out in Step~1.

Even more generally, we may choose to carry out the two-step procedure being referred to not over
the standard parameter space, but over an augmented parameter space that includes all unknown
variables on which the generation of the observed data set $x$ has been assumed to depend (but this
is a slightly more technical issue).

In spite of what has just been assumed, the reader should not get completely carried away with the
idea that we will always be looking to conduct a \pagebreak statistical analysis that is divided
over regions of the parameter space. This is because, in Step 1 of the approach just put forward,
the number of regions that the parameter space is divided into may be just one, i.e.\ this step
would consist of using a single method of inference to construct the post-data distribution of the
parameter $\theta$ over the entire parameter space.

\vspace{2ex}
\section{Fiducial inference}
\label{sec3}

So far into a sensible discussion about statistical inference and the fiducial argument has not yet
even been mentioned, let us immediately put that right!

Fiducial inference together with Bayesian inference can be viewed as the two principle (but not
only) inferential engines that drive the general IOI approach discussed in the previous section.
To begin with, we need to remember that, since we are applying the philosophy of ``no analogy, no
method'', we need to clarify what kind of analogy is required to be able to use fiducial inference.
Let us do that with regard to a classic example of fiducial inference.

Let us suppose that we wish to make inferences about the mean $\mu$ of a normal distribution that
has a known variance $\sigma^2$ on the basis of a sample $x$ of size $n$ drawn from the
distribution concerned.
Given that the sample mean $\bar{x}$ is a sufficient statistic for $\mu$, what we call the primary
random variable $\Gamma$ is naturally defined in this case as being:
\vspace{0.5ex}
\begin{equation}
\label{equ1}
\Gamma = \frac{\bar{x} - \mu}{\sigma/\sqrt{n}}
\vspace{0.5ex}
\end{equation}
We should clarify that it is convenient to imagine that the value of this variable is generated
directly (e.g.\ from a machine that generates random numbers) and that the sample mean $\bar{x}$ is
determined from this value of $\Gamma$ by simply rearranging equation~(\ref{equ1}).
Notice that the variable $\Gamma$ clearly had a standard normal distribution before the observed
sample mean $\bar{x}$ was generated.

The analogy we need to make in using the fiducial argument is with the general argument that if the
outcome of a standard physical experiment (e.g.\ randomly drawing a ball from an urn of balls) is
hidden from us and we receive a specific information packet after the experiment has taken place,
then we may still claim that the probability of any given outcome of the experiment is the same as
it was before the experiment took place if we consider that the information packet tells us nothing
about what \pagebreak the outcome of the experiment might have been.

In trying to apply this analogy in the current context, the outcome of the experiment would be
associated with the realised value of $\Gamma$ and receiving the information packet would be
associated with observing the sample mean $\bar{x}$.
However, if we were able to place a probability distribution over $\mu$ that adequately represents
what we knew about $\mu$ before the data were observed, then this analogy is unlikely to be
satisfactory as, in general, observing $\bar{x}$ will change in some way what we would conclude
about our uncertainty regarding the true value of $\Gamma$.
Applying the ``no analogy no method'' rule would therefore block the use of the fiducial argument
in this case.

On the other hand, if we had known nothing or very little about $\mu$ before the data were
observed, then it would not be a surprise if the analogy in question was regarded as being very
acceptable, and therefore we would be allowed to proceed in making inferences about $\mu$ by using
the fiducial argument.
Just to clarify, using this argument amounts to assuming that $\Gamma$ has the same distribution
that it had before the data $x$ were observed, i.e.\ it follows a standard normal distribution.
This directly implies, by rearranging equation~(\ref{equ1}), that the fiducial or post-data
distribution of $\mu$ is given by the expression:
\vspace{0.5ex}
\[
\mu\,|\,\sigma^2,x \sim \mbox{N}(\bar{x}, \sigma^2/n)
\vspace{1ex}
\]

Fiducial inference is important because:

\begin{enumerate}

\item It can be argued that the most common situation in which we may find ourselves with regard to
our pre-data knowledge about a parameter of interest is one in which the amount of such knowledge
is very limited, and in such a situation, fiducial inference often works very well.

\item The Bayesian approach to inference often falls down in this type of situation as it is
difficult to express very limited pre-data knowledge about a parameter in the form of a probability
distribution over the parameter, or to be more precise, difficult to do this in a way that means
that making the Bayes' analogy discussed in Section~\ref{sec1} is acceptable.

\end{enumerate}

The concept of fiducial inference has been developed in a way that is consistent with the
interpretation of this type of inference that has just been given in Bowater~(2017b, 2018a, 2019b).
In the last of these three papers, the resulting theory was called \emph{organic fiducial
inference}. Note that in the general case, a primary random variable (i.e.\ the variable $\Gamma$
in the example just outlined) together with the usual model parameters $\theta$ may form the
dimensions of what was referred to in the last section as an augmented parameter space.

\vspace{2ex}
\section{Multiparameter problems}

One of the biggest drawbacks of fiducial inference has been that it has proved difficult to extend
this type of inference (in both a mathematical and philosophical sense) to multiparameter problems
(i.e.\ problems where more than one parameter is unknown).
Also, this can be viewed as being a general difficulty for non-Bayesian methods of inference that,
nevertheless, may work well in facilitating the construction of a post-data distribution for a
single parameter.
On the other hand, it can be argued that the Bayesian method has a natural capacity to tackle
multiparameter problems.
However, the big assumption needs to be made, of course, that a sensible joint prior distribution
of the parameters can be constructed.

The first step of a convenient strategy for overcoming the general difficulty being referred to is
to determine the set of full conditional post-data densities for the model parameters $\theta$,
i.e.\ the set of full conditional densities:
\begin{equation}
\label{equ2}
p(\theta_j\,|\,\theta_{-j},x)\ \ \ \mbox{for $j = 1,2,\ldots,k$}
\end{equation}
(where $\theta_{-j}$ denotes all the parameters in $\theta$ except for $\theta_j$) with it
understood that any given density function in this set is determined using whatever method of
inference (i.e.\ Bayesian, fiducial or other type of inference) is appropriate for that particular
task.

As alluded to in Section~\ref{sec2}, if these full conditional densities are compatible, then under
a mild condition, they determine a unique joint post-data density for the parameters $\theta$.
This joint post-data density can sometimes be determined using analytical methods.
On the other hand, if the full conditional densities in equation~(\ref{equ2}) are not consistent
with any joint distribution of the parameters $\theta$ (in other words they are incompatible) then
very often they will be approximately compatible, meaning that a joint density function of the
parameters $\theta$ can be found that has full conditional densities that closely approximate those
given in equation~(\ref{equ2}). A way of determining the joint density of the parameters $\theta$
that best fits this criterion is by using a Gibbs sampler based on the full conditional densities
in equation~(\ref{equ2}) with an appropriately chosen scanning order of the
parame-{\linebreak}ters $\theta$. (Note that the limiting distribution of a Gibbs sampler is
affected by the scanning order of the parameters if the full conditional densities on which it is
based are not compatible).

This is the general approach for dealing with multiparameter problems when using integrated organic
inference (IOI). This approach is discussed in detail in both Bowater~(2018a) and Bowater~(2020).

\vspace{2ex}
\section{Bispatial inference}

To give another example of a method of inference that may be incorporated into the general IOI
framework outlined in Section~\ref{sec2}, let us consider bispatial inference. Here we will
concentrate on the case in which bispatial inference can be essentially viewed as just being a way
of interpreting one-sided P values, even though in fact, as explained in Bowater~(2019c), it can
regarded as being a more general type of inference (or at least general enough to warrant its
grandiose name!).

The problem of inference that we will examine will be the simple problem of inference considered in
Section~\ref{sec3}, i.e.\ the problem of making inferences about a normal mean $\mu$, and it will
be assumed that we wish to determine the post-data probability that $\mu$ is less than or equal to
a given non-negative constant $\varepsilon$, i.e.\ the probability $P(\mu \leq \varepsilon\,|\,x)$.
To be able to apply the method of bispatial inference to this problem we need to make an analogy
since as we know: no analogy, no method!

Let us consider therefore making an analogy with a situation where an experiment has been conducted
to test whether an individual does or does not have ESP (Extra\hspace{0.05em}-Sensory Perception).
The result of the experiment is a non-negative score. If the individual does indeed have ESP then
we would expect this score to be higher rather than lower. Given the observed score $y$, we
calculate the one-sided P value for the null hypothesis that the individual does not have ESP\@.
Let this value be denoted as $P_0$.
We now attempt to assess the proportion of times that the observed score would be greater than $y$
if we repeated this experiment. Let this proportion be denoted as $\mathcal{P}$.
We can appreciate that if we were able to establish that the proportion $\mathcal{P}$ was greater
than the P value $P_0$, then clearly we would have to conclude that the individual does indeed
have~ESP.

Of course, most people would have a very high degree of pre-data belief that the individual does
not have ESP\@. However, if we imagine consecutive scenarios in which $P_0$ turns out to be smaller
and smaller, then, at least after a certain point, we would surely feel that it is more and more
likely that the proportion $\mathcal{P}$ would be greater than $P_0$, and therefore gradually more
likely that the individual does indeed have ESP\@. This is of course under the assumption that we
have complete confidence in the integrity of the experiment.

In making an analogy between this situation and the original example, let us assume that in the
original example, the one-sided P value would be the one-sided P value for the null hypothesis that
$\mu \leq \varepsilon$, i.e.\ the quantity:
\vspace{1ex}
\begin{equation}
\label{equ3}
1 - \Phi \left(\frac{\bar{x} - \varepsilon}{\sigma/\sqrt{n}} \right)
\vspace{1ex}
\end{equation}
where $\Phi(\hspace{0.05em})$ represents the (cumulative) standard normal distribution function.
For this analogy to be a good analogy we would require that:

\begin{enumerate}

\item There was a substantial degree of pre-data belief that $\mu \leq \varepsilon$. This would
usually be the case given the range of values of $\mu$ that are consistent with the null hypothesis
being true. Nevertheless, to avoid the temptation of using fiducial inference to determine the
post-data probability of interest $P(\mu \leq \varepsilon\,|\,x)$, let us assume that $\varepsilon$
is a small value and there was a substantial degree of pre-data belief that $\mu$ lay in the
interval $[-\varepsilon, \varepsilon]$, i.e.\ a substantial degree of pre-data belief that $\mu$
was equal or close to zero. This would have practical relevance of course if $\mu=0$ indicated, for
example, the absence of a treatment effect.
\item The observed P value was small, at the very least, less than 0.1.

\end{enumerate}

Given the analogy we are making, it should be clear how we would therefore use the P value in
equation~(\ref{equ3}) to determine our post-data probability for the event that $\mu \leq
\varepsilon$. In summary, this probability is assessed in quite a direct manner after having taken
into account the size of the P value in question and pre-data knowledge about the parameter $\mu$.
Note that to determine the probability of interest $P(\mu \leq \varepsilon\,|\,x)$, we need, of
course, to make another analogy, namely, the type of analogy between our confidence (after the data
have been observed) that $\mu \leq \varepsilon$ and our confidence in a given outcome of a
stan-{\linebreak}dard physical experiment occurring that was discussed in Section~\ref{sec4}.

\pagebreak
The determination of the post-data probabilities that $\mu$ lies in the two regions of the real
line of concern here, i.e.\ the regions where $\mu \leq \varepsilon$ and where $\mu > \varepsilon$
would be considered as constituting Step~2 of the general framework given in Section~\ref{sec2}.
To carry out Step~1 of this framework, i.e.\ to determine the distribution of $\mu$ within each of
these regions, we could, for example, use organic fiducial inference.

This combination of bispatial inference and organic fiducial inference was described in detail in
Bowater~(2019c) and was illustrated further in Bowater~(2020). The development of simply the method
of bispatial inference itself was presented in both Bowater and Guzm\'an-Pantoja~(2019a) and
Bowater~(2019c).

\vspace{2ex}
\section{Future challenges}

To conclude this guide to integrated organic inference (IOI), let us list some priorities with
regard to the future development of this inferential framework.

\begin{enumerate}

\item Extending this framework to the case where the underlying true model is unknown.

\item Evaluating what other methods of inference could be incorporated into the IOI framework. At
the moment it would appear that along with the three methods of inference discussed in this guide
(i.e.\ Bayesian, fiducial and bispatial inference) the only additional methods that we would wish
to include in this framework would be very simple and intuitive methods of inference that may be of
service in certain special cases. However, the door should never be closed on the possible
existence of new methods, and of course their accompanying analogies, that could be used to draw
inferences about model parameters in Step~1 or Step~2 of the general framework outlined in
Section~\ref{sec2}.

\item Developing further or possibly adjusting the methods of inference that are currently
incorporated into the IOI framework, especially, given the lack of general attention they have
received, organic fiducial inference and bispatial inference.

\item Looking for ways of improving the computational side of the IOI framework. While it has been
shown how this framework can be easily applied to a variety of multiparameter problems (see
references), there are important issues that need to be overcome for this framework to have
computational demands that, in general, are comparable with, say, Bayesian inference.
However, it is of course preferable to have an approximate solution to an inferential problem that
is based on sound reasoning, rather than an exact solution that is based on flawed reasoning.

\end{enumerate}

\vspace{3.5ex}
\noindent
\textbf{References}

\vspace{2.25ex}
\noindent
Papers are listed in the order in which they were written. All papers (even the papers published
in arXiv.org) are in their final form. Free versions of these papers are available from the
author's website.

\begin{description}

\item[] Bowater, R. J. (2017a).\ A formulation of the concept of probability based on the use of
experimental devices.\ \emph{Communications in Statistics:\ Theory and Methods}, \textbf{46},
4774--4790.

\item[] Bowater, R. J. (2017b).\ A defence of subjective fiducial inference.\ \emph{AStA Advances
in Statistical Analysis}, \textbf{101}, 177--197.

\item[] Bowater, R. J. and Guzm\'an-Pantoja, L. E. (2019a).\ Bayesian, classical and hybrid methods
of inference when one parameter value is special.\ \emph{Journal of Applied Statistics},
\textbf{46}, 1417--1437.

\item[] Bowater, R. J. (2018a).\ Multivariate subjective fiducial inference.\ \emph{arXiv.org
(Cornell University), Statistics}, arXiv:1804.09804.

\item[] Bowater, R. J. (2018b).\ On a generalised form of subjective probability.\ \emph{arXiv.org
(Cornell University), Statistics}, arXiv:1810.10972.

\item[] Bowater, R. J. (2019b).\ Organic fiducial inference.\ \emph{arXiv.org (Cornell University),
Sta\-tis\-tics}, arXiv:1901.08589.

\item[] Bowater, R. J. (2019c).\ Sharp hypotheses and bispatial inference.\ \emph{arXiv.org
(Cornell University), Statistics}, arXiv:1911.09049.

\item[] Bowater, R. J. (2020).\ Integrated organic inference (IOI):\ a reconciliation of
statistical paradigms.\ \emph{arXiv.org (Cornell University), Statistics}, arXiv:2002.07966.

\end{description}

\end{document}